\documentclass[a4paper,12pt]{article}
\usepackage{amsmath}
\usepackage{amsthm}
\usepackage{amsfonts}
\usepackage{color}

\usepackage[T1]{fontenc}
\usepackage[latin9]{inputenc}
\usepackage[a4paper]{geometry}
\usepackage{graphicx}
\usepackage{psfrag}
\usepackage{babel}
\usepackage{titling}
\predate{}
\postdate{}

\newcommand{\bm}[1]{\mbox{\boldmath $#1$}}
\newcommand{\widesim}[2][1.5]{
 \mathrel{\overset{#2}{\scalebox{#1}[1]{$\sim$}}}
 }

\begin{document}

\def \k   {{\kappa}}
\def \X  {{\bf X}}
\def \Y  {{\bf Y}}
\def \g   {{\bf g}}
\def \B  {{\cal B}}
\def \D  {{\cal D}}
\def \S  {{\cal S}}
\def \H  {{\bf H}}
\def \v   {{\hbox{v}}}
\def \x  {{\bf x}}
\def \s   {{\bf s}}
\def \bt   {{\bm \theta}}
\def \y  {{\bf y}}

\vskip.5in  
\title{ Interpretable, predictive spatio-temporal models via enhanced Pairwise Directions Estimation}

\author{
Heng-Hui Lue\\
Department of Statistics, \\
Tunghai University\\
 Taichung, Taiwan\\
\textit{hhlue@thu.edu.tw}
\and ShengLi Tzeng\\
Department of Applied Mathematics,\\
National Sun Yat-sen University, \\
Kaohsiung, Taiwan\\
\textit{slt.cmu@gmail.com}}
\date{}
\maketitle

\noindent
{\bf Summary}

\noindent
This article concerns the predictive modeling for spatio-temporal data as well as model interpretation using data information in space and time.
We develop a novel approach based on supervised dimension reduction for such data in order to capture nonlinear mean structures without requiring a 
prespecified parametric model.
In addition to prediction as a common interest, this approach emphasizes the exploration of geometric information from the data.
The method of Pairwise Directions Estimation (PDE; Lue, 2019) is implemented  in our approach as a data-driven function searching for spatial patterns 
and temporal trends.
The benefit of using geometric information from the method of PDE is highlighted, which aids effectively in exploring data  structures.
We further enhance PDE, referring to it as PDE+,  by incorporating kriging to estimate the random effects not explained in the mean functions.
Our proposal can not only increase prediction accuracy, but also improve the interpretation for modeling.
Two simulation examples are conducted and comparisons are made with four existing methods.
The results demonstrate that the proposed PDE+ method is very useful for exploring and interpreting the patterns and trends for spatio-temporal data.
Illustrative applications to two real datasets are also presented.

\vskip.3in

\noindent
{\bf Keywords}
\noindent

\noindent
covariates, dimension reduction, kriging, semi-parametric models, visualization

\section{Introduction}

Complicated phenomena in spatio-temporal data raise large challenges remaining to be overcome even in today's new era of computing.
Kriging, as a method widely used in modeling  such data, typically contains two components; namely, a stationary Gaussian process and a mean function.
Interestingly, most of kriging approaches assume a very simple structure for the mean function.
For example, data at $n$ locations over $T$ time points are assumed to be observed according to 
\begin{align*}
  y(\s,t)&= \mu(\s,t)+\eta(\s,t)+\epsilon(\s,t),
\end{align*}
where $\mu(\s,t)$ is a mean function, and $\eta(\s,t)$ is a zero-mean Gaussian process with a covariance function 
$C(\s-\s^*,t-t^*) \equiv \mbox{cov}({\eta(\s,t), \eta(\s^*,t^*)})$.
Usual choice of $\mu(\s,t)$ is either a linear combination of \lq\lq known'' covariates (i.e. universal kriging), or a constant not varying over space or time 
(i.e. ordinary kriging).
It results in most analyses focusing on the covariance functions.

Martin and Simpson (2005) showed that ordinary kriging can obtain a poor prediction under the presence of strong trends.
To explore the important spatial structures for spatio-temporal data, the renowned empirical orthogonal function (EOF) analysis is a commonly used method, 
which is based on principal component analysis; see Cressie and Wikle (2011) and Demsar et al. (2013) for review.
Letting 
\begin{align}
   z(\s,t)=\mu(\s,t)+\eta(\s,t), \label{mean_y}
\end{align}
EOFs formulate the spatio-temporal data approximately in the summation form as 
\begin{align}
  z(\s,t) \approx \sum_{k=1}^K{\mbox{PC}}_k(t)\mbox{EOF}_k(\s), \label{eof}
\end{align}
 (see, Braud and Obled, 1991 and Thorson et al., 2020).
This linear combination from inner products of temporal and spatial functions in EOFs has close connection to many dimension reduction approaches for 
geostatistics in the literature.
From the reduced rank perspective, a low-rank model for spatial data at a particular time point is considered in a form similar to (\ref{eof}) as 
\begin{equation}
  z(\s) \approx \sum_{k=1}^K \alpha_k \beta_k(\s), \label{low-rank}
\end{equation}
where $\alpha_k$'s are unknown scalars and $\beta_k(\s)$ is either a fully known function or a parametric basis function depending on a few parameters 
(e.g., Cressie and Johannesson, 2008; Banerjee et. al., 2008 and Wikle, 2010).
When several time points are involved, $\alpha_k$ in (\ref{low-rank}) can be further assumed as a time-varying random variables $\alpha_k(t)$ (see, 
Cressie et. al., 2010; Fass\`o and Cameletti, 2010; Wang and Huang, 2017 and Tzeng and Huang, 2018).
Some of these works assumed $(\alpha_1(t),\ldots,\alpha_K(t))^\prime$ to be independent and identically distributed (i.i.d.) over time, while others linked 
$\alpha_k(t-1)$ and $\alpha_k(t)$ for equal time spaces in an autoregressive way.
The \texttt{SpTimer} package developed by Bakar and Sahu (2015) utilizing the low-rank framework is implemented to hierarchical Bayesian modeling for 
space-time data.
In contrast to stochastic $\alpha_k(t)$'s and parametric $\beta_k(\s)$'s above, additive models provide a nonparametric perspective on (\ref{low-rank}).
They used spline functions to represent $\alpha_k(t)$ and/or $\beta_k(\s)$ (e.g., Sharples and Hutchinson, 2005 and  Lee and Durb\'an, 2011).
The \texttt{mgcv} package based on Woods(2017) is a convenient tool for additive models.

As a motivating example for illustrating the intriguing discovery, a data of average temperatures for each day of the year at 35 weather stations in Canada, 
shown in Figures 3(A) and 3(B), are considered.
From Figure 3(A), a concave down pattern for daily temperatures is clear observed.
It seems to imply that a strong geometric information structure in the data.
Two immediate questions are raised:  are there any important time-related shapes hidden in those curves?
Does the variation of curves relate to spatial patterns across the locations of stations?
These questions stimulated us to investigate whether the inner products of spatial and temporal functions aid in exploring and interpreting the data.
Our primary aim of the study is to gain the understanding of how temperature cycles at stations vary with location and/or with time.
In general, this aim is not easy to accomplish.
The aforementioned dimension reduction approaches turn the problem more tractable through different simplifications.
For example, the EOFs impose the orthogonality on estimated components, the low-rank methods assume certain parametric function forms, and additive 
models require a set of pre-specified smooth functions.
In contrast, we relax those simplifications through an extension of supervised dimension reduction approach into the analysis without requiring pre-specified 
parametric functions.

Instead of directly taking $z(\s,t)$ in (\ref{mean_y}), we assume $z(\s,t)=\sum_{k=1}^K \alpha_k(t) \beta_k(\s)+\eta(\s,t)$ with a substantially different mean 
distinct from typical kriging.
The Pairwise Directions Estimation (PDE; Lue, 2019) is incorporated in our proposed method to estimate the $\sum_{k=1}^K \alpha_k(t) \beta_k(\s)$ in a 
data-driven way.
In this article, we focus on explicitly finding both $\alpha_k(t)$ and $\beta_k(\s)$ in order to detect potentially sophisticated mean structures in the 
spatio-temporal data.
Despite recent flourish of research on supervised dimension reduction (e.g., Li, et al., 2003; Lue, 2019 and Coudret et al., 2014), such time relevant 
high-dimensional data remains to be challenging.
We shall propose a novel approach to adaptively capture important spatial structures and temporal patterns for spatio-temporal data.
The numerical results and real applications show that our proposed method not only gives explainable mean trends, but may also produce more accurate 
model prediction.

The remainder of the paper is organized as follows. Section 2 introduces our approach.
 We use two simulation examples to illustrate and evaluate our proposed method in Section 3.
Section 4 applies our method to two real datasets, and Section 5 concludes with discussion on future work.

\section{Proposed Method}
Consider a sequence of processes, $\{z(\s,t):\s\in \D\}$ for $t=1,\ldots, T$, defined on a $d$-dimensional spatial domain $\D \subset \Re^d$ with $T\ge 1$.
The processes are assumed to have a stationary spatio-temporal covariance function $\hbox{cov}(z(\s,t),z(\s^*,t^*))$.
The spatio-temporal random effects model of $y(\s,t)$ with a measurement error $\varepsilon$ is considered in this article as 
\begin{align}
y(\s,t)&=z(\s,t)+\varepsilon(\s,t),  \label{m1} \\ 
z(\s,t)&=\bm{w}(t)\bm{f}(\s)+u(\s,t),  \label{m2} 
\end{align} 
for $t=1,\ldots,T$, where $\bm{f}(\s)=(f_1(\s),\ldots,f_\k(\s))'$, $\bm{w}(t)=(w_1(t),\ldots,w_\k(t))$, $u(\s,t)$ is a zero-mean random effect with a covariance 
function $C(\s-\s^*,t-t^*)=\mbox{cov}({u(\s,t), u(\s^*,t^*)})$ for any pair of $(\s,t)$ and $(\s^*,t^*)$, 
and $\varepsilon(\s,t)$ $\sim N(0, \sigma_\varepsilon^2(t))$ is an additive white-noise uncorrelated with $z(\s,t)$.
We assume that $u(\cdot,t)$ and $\varepsilon(\cdot,t)$ are mutually uncorrelated.
More specifically, $f_{j}(\s_i)$ stands for $f_j( \bt_j'\x(\s_i) )$, where $\x(\s_i)$ is a spatial-only covariate vector, and we can represent the 
inner-product term in (\ref{m2}) in terms of the summation as 
\begin{equation}
  \bm{w}(t)\bm{f}(\s)=\sum_{j=1}^\k w_j(t) f_j(\s). \label{m3}
\end{equation} 
Here $w_j(t), j=1,\ldots,\k$, is an unknown basis function changing gradually with time $t$, and $f_j(\s)$ is an unknown coefficient function standing for a 
certain spatial pattern.
The function $f_j(\s)$ depends on a $p\times 1$ vector $\x(\s)$ through an unknown vector $\bt_j$ of weights.
We refer to $\bt_j'\x(\s)$ as a variate.
Hence, determining $f_j(\s)$ amounts to finding both $f_j$ and $\bt_j$, which is a typical  theme in supervised  dimension reduction via constructing new 
variate $\bt_j'\x(\s)$ instead of using $\x(\s)$ directly.

For covariates $\x(\s)$ over $\Re^d$, $(s_1, \ldots, s_d, s_1^2, \ldots, s_d^2)'$ are considered in our simulation examples 
and real data applications.
We certainly can incorporate more elaborate and domain-specific covariates into $\x(\s)$, which may be useful to comprehend the 
spatio-temporal phenomenon.

Note that in the semi-parametric models of (\ref{m1}) and (\ref{m2}), only $\x(\s)$ and $y(\s,t)$ are observed, but other quantities are kept unknown.
The aim of this study is to predict the process $z(\s,t)$ by reconstructing those unknown terms $\bm{w}(t)$, $\bm{f}(\s)$ and $u(\s,t)$ based on the 
observed data $\x(\s_i)$ and $y(\s_i,t)$, $i=1,\ldots,n$; $t=1,\ldots, T$, as parsimonious as possible.
To estimate those terms in (\ref{m2}), we begin with finding $\bm{w}(t)$ and $\bm{f}(\s)$ via PDE proposed by Lue (2019), and then estimate $u(\s, t)$ 
by applying the kriging method for spatio-temporal data (e.g., Sherman, 2011, and Cressie and Wikle, 2015).
We briefly introduce the two key building blocks, PDE and kriging, and then propose our estimation method.

\subsection {Pairwise Directions Estimation}
Suppose that $y(\s,t)$ at $n$ distinct locations, $\s_1,\ldots,\s_n \in \D$, for $t=1,\ldots,T$ with a covariate vector $\x(\s)$ are available.
The PDE method was originally designed to predict the following curve data model proposed by Li {\it{et al}}. (2003) 
\begin{align}
y(\s,t)=\sum_{j=1}^\k \bm{\phi}_j(t) g_j(\bt_j'\x(\s))+e(\s,t),   \label{cdm} 
\end{align} 
for $t=1,\ldots,T$.
The original setting for time point $t$ is not necessarily equal-spaced, but for simplicity we assume them to be equal-spaced.
We allow $\bm{\phi}_j$, $g_j$ and $\bt_j$ to be determined by the data with the number $\k$ as small as possible.
The summation term in (\ref{cdm}) is used to approximate the inner-product term in (\ref{m2}).

In order to describe $y(\s,t)$, we need to estimate a $T\times 1$ vector $\bm{\phi}_j$, a $p\times 1$ vector $\bt_j$ and a link function $g_j$, which all 
are totally unknown.
The estimation strategy is to find $\bt_j$ first, and then estimate $\bm{\phi}_j$ and $g_j$ in an iterative way.
For ease illustration, we denote $\y(\cdot,t)=(y(\s_1,t),\ldots,y(\s_n,t))'$, $\y(\s,\cdot)=(y(\s,1),\ldots,y(\s,T))'$ and ${\bf Y}=(\y(\cdot,1),\ldots,\y(\cdot,T))$.
First of all, we introduce two methods to obtain an initial estimate $\hat{\bt}_j$ of $\bt_j, j =1,\ldots,\k$, for efficiently implementing the PDE method.
One choice is based on mrSIR method of Lue (2009).
The initial $\hat \bt_j$ can be found by solving the eigenvalue decomposition of ${\bf \bar{\Psi}}$ with respect to ${\bf \Sigma}_{\x}$, 
\begin{align}
 {\bf \bar{\Psi}}\,\bt_j=\rho_j {\bf \Sigma}_{\x}\, \bt_j,  \label{mrsir}
\end{align}
where ${\bf \bar{\Psi}}=\sum_{t=1}^T (\bar \gamma_t/\bar\gamma_{.}){\bf \Psi}_t$, ${\bf \Psi}_t=\hbox{var}\{E(\x(\s)|\y(\cdot,t))\}$, $\bar \gamma_t$ 
is the proportion of nonzero eigenvalues for the eigenvalue decomposition of ${\bf \Psi}_t$ with respect to ${\bf \Sigma}_\x=\hbox{var}(\x(\s))$, and 
$\bar\gamma_{.}=\sum_t \bar\gamma_t$.

Another choice is derived from pe-mrPHD method of Lue (2010) which is delineated as follows: 
It begins by forming the $T$ principal components $\tilde \y_t$ based on ${\bf Y}$ associated with eigenvalues $\lambda_t$, $t=1,\ldots,T$.
Construct the covariance matrix ${\bf \Sigma}_{\tilde{\y}_t\x\x}=E\{(\tilde{\y}_t-E\tilde{\y}_t)(\x(\s)-E\x(\s))(\x(\s)-E\x(\s))'\}$, and then let ${\bf \Phi}_t\,\hbox{diag}(\lambda_{1t},\ldots,\lambda_{pt}){\bf \Phi}_t^\prime$ be its eigenvalue 
decomposition, where ${\bf \Phi}_t=(\bm {\psi}_{1t},\ldots,\bm {\psi}_{pt})$ 
and $\bm {\psi}_{it}$ is the eigenvector corresponding to eigenvalue $\lambda_{it}$ for $ i=1,\ldots,p$.
Define a positive eigenvalue version for ${\bf \Sigma}_{\tilde{\y}_t\x\x}$ as $\H_t={\bf \Phi}_t\,\hbox{diag}(|\lambda_{1t}|,\ldots,|\lambda_{pt}|){\bf \Phi}_t^\prime$, 
and let their weighted average be $\bar{\H}=\sum_{t=1}^T (\lambda_t/ \lambda_{.}) \H_t$, where $\lambda_{.}=\sum_{t=1}^T \lambda_t$.
Then pe-mrPHD conducts the eigenvalue decomposition of $\bar{\H}$ with respect to ${\bf \Sigma}_{\x}$, 
\begin{align}
 \bar{\H}\, \bt_j =\rho_j {\bf \Sigma}_{\x}\, \bt_j,  \label{mrphd}
\end{align} 
to obtain the initial value of $\hat \bt_j$.
Note that equations in (\ref{mrsir}) and (\ref{mrphd}), operations of $E(\cdot)$ and $\hbox{var}(\cdot)$ are performed through the sample version in practice.

After initialization of $\bt_j$, we utilize an adaptive estimation of MAVE (Xia {\it{et al}}., 2002) for a single-index case to obtain initial estimate of $\bm{\phi}_j$ 
by solving the minimization problem:
\begin{align}
   \hbox{min}_{\tilde{d}_{j_1},\tilde{d}_{j_2},\bm{\phi}_j} \sum_{\ell=1}^n\sum_{i=1}^n \bigg(\bt_j'\x(\s_i)-\{\tilde d_{j_1\ell}+\tilde d_{j_2\ell}\,
                                   \bm{\phi}_j'(\y(\s_i,\cdot)-\y(\s_\ell,\cdot))\} \bigg)^2 \delta_{i\ell},  \label{mave.0}
\end{align} 
where $\delta_{i\ell}=K_{h_{\bf Y}}\{\bm{\phi}_j'(\y(\s_i,\cdot)-\y(\s_\ell,\cdot))\}/\sum_{i=1}^n K_{h_{\bf Y}}\{\bm{\phi}_j'(\y(\s_i,\cdot)-\y(\s_\ell,\cdot))\}$, $K$ 
is a kernel function and $h_{\bf Y}$ is the bandwidth of ${\bf Y}$.

The iterative algorithm for estimating $\bm{\phi}_j$, $\bt_j$ and $g_j$ for $j=1,\ldots,\k$  in (\ref{cdm}) via PDE proceeds in the following steps:
\vskip .1in
\begin{itemize}
\item [1.]
Choose initial estimates $\hat \bt_{j(0)}$, $j =1,\ldots,\k$, from (\ref{mrsir}) and/or (\ref{mrphd}), and then get $\tilde{\bm{\phi}}_{j(0)}$ as the minimizer 
of $\bm{\phi}_j$ in (\ref{mave.0}).
Compute the estimated basis functions $\hat{\bm{\phi}}_{j(0)}=\hat {\bf \Sigma}_{\bf Y}\tilde{\bm{\phi}}_{j(0)}$, where $\hat {\bf \Sigma}_{\bf Y}$ is the 
sample covariance of ${\bf Y}$ (see, Lemma of Lue (2019)).

\item [2.]
At the $\tau$-th iteration, fit a linear regression model to $\y(\s_i,\cdot)$ against $\{\hat{\bm{\phi}}_{j(\tau-1)}\}_{j=1}^\k$, i.e. assuming 
\begin{align}
          \y(\s_i,\cdot)=g_{1i} \hat{\bm{\phi}}_{1(\tau-1)}+\cdots+g_{\k i}\hat{\bm{\phi}}_{\k(\tau-1)}+\bm{e}(\s_i,\cdot),   \label{linear.fit}
\end{align} 
for $i=1,\ldots,n$, to obtain the estimated coefficients $\hat g_{ji(\tau)}$, $j =1,\ldots,\k$, where the $(\tau)$ in the subscript denotes the iteration number.

\item [3.]
Obtain the updated estimate $\tilde{\bm{\phi}}_{j(\tau)}$ from (\ref{mave.0}) by replacing $\bt_j'\x(\s_i)$ with $\hat g_{ji(\tau)}$ and then compute the 
updated basis functions $\hat{\bm{\phi}}_{j(\tau)}=\hat{\bf \Sigma}_{{\bf Y}}\tilde{\bm{\phi}}_{j(\tau)}$, $j =1,\ldots,\k$.

\item [4.]
Repeat steps 2 and 3 until $||\hat{\bm{\phi}}_{j(\tau)}-\hat{\bm{\phi}}_{j(\tau-1)}||<\Delta, j =1,\ldots,\k$, for some given tolerance value $\Delta$ 
(e.g., $\Delta=0.001$).

\item [5.]
Use the final estimate $\hat{\bm{\phi}}_{j}$ in step 4 to obtain the final estimates $\hat \bt_{j}$ and $\hat g_{j}$, $j =1,\ldots,\k$, by solving the 
minimization problem:
\[   \hbox{min}_{d_{j_1},d_{j_2}, \bt_j} \sum_{\ell=1}^n\sum_{i=1}^n \bigg(\hat{\bm{\phi}}_j'\y(\s_i,\cdot)-\{d_{j_1\ell} +d_{j_2\ell}\, \bt_j'
                                           (\x(\s_i)-\x(\s_\ell))\}\bigg)^2 \delta_{i\ell}, 
\]
where  $\delta_{i\ell}=K_{h_\x}\{\bt_j'(\x(\s_i)-\x(\s_\ell))\}/\sum_{i=1}^n K_{h_\x}\{\bt_j'(\x(\s_i)-\x(\s_\ell))\}$ and $h_\x$ is the bandwidth of $\x(\s)$.
\end{itemize}

\subsection {Incorporation of kriging}
Since PDE does not take spatio-temporal covariance function into account, we incorporate kriging to estimate the random effects for model prediction.
Under the assumption of stationary Gaussian process, the minimum mean squared error linear prediction for $u(\s,t)$ at $(\s_0,t_0)$ of interest can 
be found by utilizing the optimality of conditional expectations under joint normality; see Cressie (2015).
The kriging method dedicates to model the covariance $C(\s-\s^*,t-t^*)=\hbox{cov}(u(\s,t),u(\s^*,t^*))$ to describe the strength of dependency 
between any pairs of variables at  $(\s,t)$ and $(\s^*,t^*)$.
Under the stationary assumption of $u(\s,t)$,  an equivalence exists between variogram $\nu(\s-\s^*,t-t^*) \equiv E(y(\s,t)-y(\s^*,t^*))$ and the covariance 
 function $C(\cdot,\cdot)$; see Cressie (2015) or  Pebesma  and  Heuvelink (2016).
We consider the isotropic product-sum model in this article with covariance in the form of 
\[  C(h_{s}, h_{t})=k_1 C_s(h_s;\bm{\xi}_1)+ k_2 C_t(h_t;\bm{\xi}_2) + k_{3}C_s (h_s;\bm{\xi}_1)C_t(h_t;\bm{\xi}_2)
\]
for any pair of space-time points $(\s,t)$ and $(\s+h_s, t+h_t)$ with valid covariance functions $C_{s}(h_{s};\bm{\xi}_1)$ and $C_{t}(h_{t};\bm{\xi}_2)$, 
respectively.

For approximation of $u(\s,t)$, we need to calculate the empirical variogram of $\{y(\s,t)\}$; namely, the sample version of $\nu(\s-\s^*,t-t^*)$.
The estimation of parameters $k_1, k_2, k_3, \bm{\xi}_1$ and $\bm{\xi}_2$ is based on least squares variogram fitting, which is the default method 
implemented in the package $\texttt{gstat}$.
After obtaining $\hat{\bm{w}}(t)\hat{\bm{f}}(\s)$ by PDE in Section 2.1, we apply kriging to estimate $u(\s,t)$ based on the residuals 
$y(\s,t)-\hat{\bm{w}}(t)\hat{\bm{f}}(\s)$ through using 
\[  E\left[u(\s_0,t_0)\bigg|\left\{y(\s_i,t)- \sum_{j=1}^\k \hat{w}_j(t)\hat{f}_{j}(\s); i=1,\ldots,n, t=1,\ldots,T \right\}\right] 
\] 
with plug-in parameters under the joint normality assumptions of $u(\cdot,\cdot)$ and $\varepsilon(\cdot,\cdot)$.
Here $\hat{w}_j(t)$ and $\hat{f}_{j}(\s)$ are replaced respectively with $\hat{\bm{\phi}}_{j}$ and $\hat g_{j}$ found by PDE in step 5 for $j=1,\ldots,\k$.
We perform one-more-step iteration to obtain the final estimate for each component in (\ref{m2}), whose algorithm is given in Section 2.4.

\subsection {Function Scaling Estimation}
The PDE captures the inner-product term, $\bm{w}(t)\bm{f}(\s)$, of nonlinear mean structures in (\ref{m2}) without requiring a prespecified parametric 
model; namely, $\hat{\bm{\phi}}_{j}$ and $\hat g_{j}$ found by PDE in step 5 are used to respectively estimate $w_j$ and $f_j$ in (\ref{m3}) for $j=1,\ldots,\k$.
In order to address the identifiability issue of functional estimation, we let the estimate $\hat{w}_j=(\hat{w}_j(1), \ldots, \hat{w}_j(T))^\prime$ be a normalized 
vector of unit norm, and then a scaling factor $\hat{b}_j(\s)$ is introduced for prediction purpose.
Under (\ref{m2}), we use $\sum_{j=1}^\k \hat{w}_j(t) \hat{b}_j(\s) \hat{f}_{j}(\s)$ to approximate $\bm{w}(t)\bm{f}(\s)$ via a linear fit.
More specifically, $\hat{b}(\s_i)=(\hat{b}_1(\s_i),\ldots,\hat{b}_\k(\s_i))'$ are the estimated coefficients obtained from the linear regression of $y(\s_i,t)$ 
against $\{\hat{f}_j(\s_i)\hat{w}_j(t)\}_{j=1}^{\k}$ for each $i=1,\ldots,n$.
Then we refer to the $j$th scaled coefficient function $\tilde{f}_{j}(\s)$ as the estimates 
$(\hat{b}_j (\s_1)\hat{f}_j(\s_1),\ldots,\hat{b}_j (\s_{n}) \hat{f}_j(\s_{n}))^\prime$, $j=1,\ldots,\k$.
Based on the scaling procedure, it would not only improve the prediction accuracy, but also give appropriate magnitude for the inner products.
In calculating $\hat{\bf z}(\s_0,\cdot)$ at a predicted location $\s_0$, the scaled coefficients $\hat{b}_j(\s_0)\hat{f}_j(\s_0), j=1,\ldots,\k$, can be obtained 
by using the average of k-nearest neighbors around $\hat{\bt}_j'\x(\s_0)$ in the observed data (e.g. k=3).

\subsection {Enhanced PDE}
We propose a new enhanced method of PDE, called PDE+, through incorporating the PDE with kriging for modeling spatio-temporal data.
In what follows, $\bm{\epsilon}(\s,\cdot)$, $\hat{\bm{\epsilon}}(\s,\cdot)$, $\bm{u}(\s,\cdot)$, and $\hat{\bm{u}}(\s,\cdot )$ are defined in a similar way as 
$\y(\s,\cdot)$ with respect to $\{y(\s,t);\; t=1,\ldots,T\}$.
The algorithm for PDE+ proceeds in the following two steps:

\vskip .1in
\begin{itemize}
\item [I.] 
Apply PDE to the data $\{\x(\s),y(\s,t)\}$ for obtaining $\hat{\bm{f}}(\s)$ and $\hat{\bm{w}}(t)$, and then get the residuals $\bm{\epsilon}(\s_i,\cdot)$ from 
a linear fit of $\y(\s_i,\cdot)$ against $\{\hat{w}_j\}_{j=1}^\k$ for $i=1,\ldots,n$.
Use kriging in Section 2.2 on $\bm{\epsilon}(\s,\cdot)$ to get an initial estimate $\hat{\bm{u}}(\s,\cdot)$ of $\bm{u}(\s,\cdot)$.

\item [II.] 
Let $\hat{\bm{\epsilon}}(\s,\cdot)$ be the difference between $\y(\s,\cdot)$ and $\hat{\bm{u}}(\s,\cdot)$ from step I, and then repeat step I once by treating 
the observed data as $\{\x(\s),\hat{\bm{\epsilon}}(\s,t)\}$.
\end{itemize}
The final predictive model of PDE+ is defined as the one in step II.
The reason for taking once iteration on PDE+ is for computational efficiency.
For prediction, we incorporate function scaling estimation procedure in Section 2.3 for measuring the prediction error based on the testing data.

\section{Numeric Results}

\subsection{Simulation Setups}
We shall apply our proposed method to dimension reduction for studying predictive modeling on spatio-temporal data through two simulated examples.
Suppose that we observe data $y(\s,t)$ at location $\s$ and time $t$.
Let the spatial-only covariate vector be $\x(\s)=(s_1,s_2,s_1^2,s_2^2)'$, where $\s=(s_1,s_2)'$, generated from given spatial  locations and $f_j(\cdot)$ 
be the coefficient function for unknown spatial structure shared across $t$.
Under (\ref{m2}), we think of $\bm{w}(t)\bm{f}(\s)$ as unknown deterministic functions.
 To make comparisons with our proposal, we consider four more existing methods.
To evaluate the performance of model prediction among various methods, we compute two versions of prediction error criteria, namely root integrated 
mean squared error (RIMSE) and rooted prediction mean squared error (RPMSE), defined by 
\begin{equation}
\begin{split}
\textrm{RIMSE}=& \frac{1}{n_{t}}\sum_{i=1}^{n_{t}}||\y(\s_{i},\cdot)-\hat{\bf z}(\s_{i},\cdot)||,  \\ 
\textrm{RPMSE}=& \left( \frac{1}{n_{t}T}\sum_{t=1}^{T}\sum_{i=1}^{n_{t}}(y(\s_{i},t)-\hat{z}(\s_{i},t))^{2} \right) ^{1/2}, 
\end{split} 
\label{pmse}
\end{equation}
where $n_t$ is the size of locations in the testing data, $\hat{z}(\s_i,t)$ denotes the estimate of $z(\s_i,t)$, and 
$\hat{\bf z}(\s_{i},\cdot)=(\hat{z}(\s_i,1), \ldots, \hat{z}(\s_i,T))'$.
The cross-validation estimates of RIMSE and RPMSE are collected over 100 replications for each example.
The four methods, referred to as naive, SpTimer, kriging and mgcv, are briefly reviewed below.

The naive method is the simplest predictor $\hat{z}(\s,t)= \frac{1}{n_{t}}\sum_{i=1}^{n_{t}}y(\s_{i},t)$, i.e., the sample mean of all locations at time $t$.
The SpTimer method uses hierarchical Bayesian modeling for space-time data. Its model is given by 
\begin{align*}
y(\s,t)&=\alpha+R(\s,t)+\varpi(\s,t)\\ 
R(\s,t)&=\rho R(\s,t-1)+\varsigma(\s,t),
\end{align*}  
where $\varsigma(\s,t)$'s are mean-zero spatial random effects independent over time with an isotropic stationary Mat{\'e}rn covariance function.
This method has been  implemented in an R package named \texttt{SpTimer}.
The kriging method typically considers the case of $\bm{w}(t)\bm{f}(\s)$ being 
a constant $\mu$.
Then ordinary kriging estimates $\mu+u(\s_0,t_0)$ by $\sum_{i=1}^n\sum_{t=1}^T p_{it}(\s_0,t_0) y(\s_i,t)$, where $\{p_{it}(\s_0,t_0)\}$ depend on 
the variogram $\nu(\cdot,\cdot)$ with the constraint $\sum_i\sum_t p_{it}(\s_0,t_0)=1$, with plugging all estimated covariance 
parameters.
Finally, the mgcv method exploits additive models with tensor products. It assumes 
\begin{equation}
 E[y(\s,t)]=q_{1}(s_{1})+q_{2}(s_{2})+q_{3}(t)+q_{4}(s_{1},s_{2})+q_{5}(s_{1},t)+q_{6}(s_{2},t)+q_{7}(s_{1},s_{2},t), \label{mgcv}
\end{equation}
where $q_k$'s are constructed by superimposing several known basis functions and its noises are assumed to follow $N(0, \sigma_y^2)$ 
independently over space and time.
With these higher-order interactions, the model tends to overfit, and hence some penalized  approach is necessary.
The famous R package \texttt{mgcv} provides convenient ways for determining an appropriate penalty.

According to the algorithms of the above methods, the default values are used to set the turning parameters.
We emphasize the accuracy of prediction as the criteria in (\ref{pmse}) and the findings of interesting pattern structures.
From simulations and practical applications in the next two sections, PDE successfully achieves the accuracy of the EDR direction estimation.

\subsection{Comparison of Simulation Results}
In each simulation run, we set $T=20$ and generate $\y(\s_i,\cdot)$ at $n$ locations which consist of a learning set with size $n_\ell (=80\%\times n)$ 
and a testing set with size $n_t (=20\%\times n)$.
The location set $\{\s_1,\ldots,\s_n\}$ is drawn from $\D=[-1,1]^2$ using simple random sampling.
We set the number of slices for mrSIR to be 10 for all runs.
We denote the absolute value of cosine angle between the estimate $\hat \bt_j$ and the true vector $\bt_j$ as $|\cos(\hat \bt_j)|$, $j=1,\ldots,\k$, for 
evaluating the accuracy of directional estimation.

\vskip0.1in
{\it Example 1}.
Consider a process, $\{z(\s,t):\s\in [-1,1]^2\}$, generated according to (\ref{m2}) with trigonometric and quadratic structures by setting 
\begin{equation}
\begin{split}
f_1(\s)&=\cos(0.5\pi ||\s-(-0.5, -0.5)'||^2),  \\ 
f_2(\s)&=\sin(0.5\pi ||\s-(0.5, 0.5)'||^2), 
\end{split} 
\label{ex1a}
\end{equation}
\begin{align}
w_{1}(t)&=(0.5\,t-5)^2,\,\,w_{2}(t)=5\sin(0.1\pi\,t),\label{ex1b}
\end{align}
for $t=1,\ldots,T$, and $u(\s,t)$ being a zero-mean random effect with the covariance 
\begin{align*}
\textrm{cov}(u(\s,t),u(\s^{*},t^{*})) = 
 \begin{cases}
    \exp(-0.5|\s-\s^{*}|) & \textrm{if}\:t=t^{*};\\
    0 & \textrm{otherwise}.
 \end{cases}
\end{align*}
Using (\ref{m1}), the data $\{y(\s,t)\}$ with $\varepsilon(\s,t) \widesim[2]{i.i.d.} \textrm{N}(0, 0.25)$ over time and space are generated.
With letting $\bt_1=(0.5,0.5,0.5,0.5)'$ and $\bt_2=(-0.5,-0.5,0.5,0.5)'$, the coefficient functions $f_j$'s can be reformulated as 
$f_1(\s)=\cos(0.5\pi (2\bt_1'\x(\s)+0.5))$ and $f_2(\s)=\sin(0.5\pi (2\bt_1'\x(\s)+0.5))$, where $\x(\s)=(s_1,s_2,s_1^2,s_2^2)'$.
Both unknown vectors, $\bt_1$ and $\bt_2$, are sufficient dimension reduction directions needed to be estimated.

To illustrate the application of PDE+ via data visualization, a single run is taken.
The data with $n=100$ are generated according to models (\ref{ex1a}) and (\ref{ex1b}).
We separately conduct the eigenvalue decomposition in (\ref{mrsir}) and (\ref{mrphd}) for the learning data to obtain initial estimates $\hat{\bt}_j$'s.
It turns out that the modified eigenvalues (0.73, 0.31, 0, 0) found by mrSIR suggest one $\x(\s)$ variate.
One significant direction is also found by pe-mrPHD with the eigenvalues (2.24, 1.83, 0.68, 0.29).
By choosing $h_Y=3, h_\x=0.5$ and taking the two initial estimates $\hat \bt_j, j=1,2$, we proceed with the proposed algorithm to dimension reduction for 
$\x(\s_i)$ and $y(\s_i,t)$, $i=1,\ldots,n$; $t=1,\ldots,T$.
After attaining the iterative convergence, two leading directions, (0.501, 0.506, 0.490, 0.502)$'$ and (-0.496, -0.496, 0.515, 0.492)$'$, for $\x(\s)$ variates 
are found along with $|\cos(\hat \bt_1)|=0.999$ and $|\cos(\hat \bt_2)|=0.999$, which are consistently close to the theoretical vectors $\bt_j, j=1,2$.
The scatterplot of the first estimated basis function, shown in Figure 1(A), reveals a noticeable  quadratic pattern.
Figure 1(B) shows a clear sine pattern for the second estimated basis function.
Figures 1(C) and 1(D) display coefficient functions which are very close to the true trigonometric patterns.
Based on our final estimates and using the testing data, the $\textrm{RIMSE}=3.445$ and $\textrm{RPMSE}=0.807$ are found by PDE+ in this single run.

In reference to the sampling performance, the results based on 100 simulated replicates according to (\ref{ex1a}) and (\ref{ex1b}) are summarized in Table 1.
It reports the mean and standard deviation for RIMSE and RPMSE obtained by all methods.
PDE+ does outperform with the smallest averaged prediction error values for this model because of its capture of more clear signal for functional patterns.
The mean of RIMSE obtained by PDE+ is about 4.517, which is about  30.9\% $(= 5.912/4.517 -1)$ improvement over kriging; while the improvement with 
respect to RPMSE is more significant about 36.2\% $(=1.618/1.188-1)$.
Not surprisingly, PDE uses locally linear information via smoothing techniques to capture the function forms of $w_j$ and $f_j$, $j=1,\ldots,\k$, which make 
the rest of process more well-suited for typical kriging models.
The effectiveness of prediction is also validated via combining the strengths of PDE and kriging into PDE+.
In contrast, SpTimer and naive methods produce considerable mean squared error in estimation, however.

\vskip0.1in
{\it Example 2}.
Generate a process, $\{z(\s,t):\s\in [-1,1]^2\}$, from (\ref{m2}) with setting 
\begin{equation}
\begin{split}
f_1(\s)&=15/\{-0.75+\exp(||\s-(-0.5, -0.5)'||^2) \},  \\ 
f_2(\s)&=1.5(-2+||\s-(0.5, 0.5)'||^2)^2, 
\end{split} 
\label{ex2a}
\end{equation}
\begin{align}
w_{1}(t)&=\arctan(0.1\pi\,t),\,\,w_{2}(t)=2\log(0.75+(0.1t-1)^2),\label{ex2b}
\end{align}
for $t=1,\ldots,T$, and $u(\s,t)$ being a zero-mean random effect with the covariance 
\begin{eqnarray*}
\textrm{cov}(u(\s,t),u(\s^{*},t^{*})) &=& 0.25\exp(-0.5||\s-\s^{*}||)\exp(-0.8||t-t^{*}||) \\
             &&  +\exp(-0.5||\s-\s^{*}||)+0.5\exp(-0.8||t-t^{*}||).
\end{eqnarray*}
We also generate $\varepsilon(\s,t) \widesim[2]{i.i.d.} \textrm{N}(0, 0.5)$ over time and space.
We need to estimate two sufficient dimension reduction vectors, $\bt_1=(0.5,0.5,0.5,0.5)'$ and $\bt_2=(-0.5,-0.5,0.5,0.5)'$, as previous mentioned.

The data with $n=150$  are generated according to models (\ref{ex2a}), (\ref{ex2b}) and (\ref{m1}) for a single run.
Using the eigenvalue decomposition in (\ref{mrsir}) and (\ref{mrphd}), the modified eigenvalues (1.50, 0.35, 0, 0) found by mrSIR suggest one $\x(\s)$ variate 
and another significant direction is  also found by pe-mrPHD with the eigenvalues (2.50, 1.93, 0.65, 0.50).
Two leading directions for $\x(\s)$ variates, ( 0.520, 0.480, 0.530, 0.467)$'$ and (-0.510, -0.488, 0.517, 0.483)$'$, are found by taking $h_Y=10, h_\x=0.5$, 
after attaining the iterative convergence.
The estimates along with $|\cos(\hat \bt_1)|=0.998$ and $|\cos(\hat \bt_2)|=0.999$ are consistently close to the theoretical vectors $\bt_j, j=1,2$.
The first two estimated basis functions, shown in Figures 2(A) and 2(B), reveal clear arctangent and quadratic patterns.
Figures 2(C) and 2(D) display estimated coefficient functions which are very close to the true views.
The $\textrm{RIMSE}=3.776$ and $\textrm{RPMSE}=1.025$ are found by PDE+ for the testing data.

The results based on 100 simulation runs according to (\ref{ex2a}) and (\ref{ex2b}) are summarized in Table 2.
As regards the predictive accuracy, PDE+ does outperform with the smallest averaged RIMSE and RPMSE compared to other methods.
The mean of RIMSE obtained by PDE+ is about 5.759, which is about  28\% $(= 7.375/5.759 -1)$ improvement over kriging; while the improvement is 
more significant about 58\% ($=2.612/1.654-1)$ with respect to RPMSE.
In contrast, the naive, SpTimer and mgcv produce considerable mean squared error in estimation, however.

\section{Applications to Real Data}
To illustrate the application of our approach to predictive modeling for empirical studies, we analyze two real datasets.
In the following datasets, the longitudes and latitudes are available for given spatial  locations.
Let the spatial-only covariate vector be $\x(\s)=(s_1,s_2,s_1^2,s_2^2)'$, where $s_1$ is the longitude and $s_2$ is the latitude.

\subsection{Canadian Weather Data}
We applied the proposed method to a dataset of average temperatures over the years 1960 to 1994 for each day of the year at 35 weather stations in Canada.
The data are available in the \texttt{fda} package on Comprehensive R Archive Network (CRAN).
The observations $\{y(\s,t)\}$ of interest are the average of temperatures measured on 35 location sites for 365 days.

The annual temperature cycle at 35 stations shown in Figure 3(A) indicates a clear concave-down pattern.
The map of observation locations is shown in Figure 3(B).
For illustration, we first standardize $\x(\s)$ for being centered and removing scale.
We randomly partition $\{\y(\s_i,\cdot); i=1,\ldots,35\}$ into a learning set with size 30 and a testing set with size 5 for further comparison analysis.
Details in one single analysis run are demonstrated.
We conduct the eigenvalue decomposition for the learning data to obtain initial estimates $\hat \bt_j$'s.
Two significant directions are found by pe-mrPHD with the eigenvalues (2.71, 2.35, 1.20, 0.81).
By choosing the bandwidths, $h_Y=3.5$ and $h_\x=1$, we proceed with the proposed algorithm to dimension reduction for $\x(\s_i)$ and $y(\s_i,t)$, 
$i=1,\ldots,30$; $t=1,\ldots,365$.
After convergence, two leading directions are (-0.109, -0.950, 0.217, -0.198)$'$ and (0.314, 0.110, 0.900, 0.281)$'$.
It shows that there exists respectively dominant effects, {\it latitude} and {\it squared longitude}, for the first two variates.

According to (\ref{m2}), both basis functions $w_j{(t)}$ and coefficient functions $f_j(\s)$ affect $z(\s,t)$ through the inner product form.
Figures 3(C) and 3(D) display the estimated  basis functions which reveal similar concave patterns.
Figure 3(C) indicates that $\hat{w}_1$ is positive and has much higher values in winter at the left and right sides of the plot but lower values in summer 
shown in the middle of the horizontal axis (Day).
Note that $\hat{f}_1$ in Figure 3(E) has negative values for a great portion of $\hat \bt_1'\x(\s)$, where large (small) values of $\hat \bt_1'\x(\s)$ represent 
southern (northern) locations due to the dominant negative latitude values in $\hat \bt_1$.
As a result, the influence of the first dominant component $\hat{w}_1(t)\hat{f}_1(\s)$ just goes consistently with the common sense that the larger latitude 
is, the lower temperature is in winter; the lower latitude is, the higher temperature is in summer.
The left end of Figure 3(F) is around Newfoundland and Labrador while the right end corresponds to Yukon and British Columbia because the positive 
squared longitude is roughly dominant to the horizontal axis $\hat \bt_2'\x(\s)$.
The shape of Figure 3(D) roughly contrasts the  temperature difference between summer and winter, due to these two seasons with the opposite signs 
of the function value for $\hat{w}_2$.
Consequently, $\hat{w}_2(t)\hat{f}_2(\s)$ shows a secondary component that the temperature difference in the middle of Canada is approximately half 
of that in the east or west coasts.
The secondary component also implies that coastal areas tend to have lower temperature in winter, compared to the continental regions at the similar 
latitude; see Figure 3(F).

In reference to the sampling performance, the results based on 100 sets of random data partitions as previously mentioned are summarized in Table 3.
Without using stronger smoothness constraints in (\ref{mgcv}) (e.g., dropping the $q_7$ term), mgcv produces a mean of 1405.61 with a standard 
error 6269.96 for RIMSE and a mean of 145.03 with a standard error 643.53 for RPMSE.
The larger values of RIMSE and RPMSE found by mgcv indicate its heavy overfitting.
It should be aware of the constraint impact on model complexity, even though mgcv provides its automatic penalty determination.
The results for mgcv in Table 3 are obtained through excluding $g_7$ term.
Not unexpectedly, PDE+ indeed captures the interesting functional patterns for $w_j(t)$ and $f_j(\s)$, $j=1,2$, useful for improving model prediction.
The PDE+ does outperform with the smallest averaged prediction error through integrating PDE for the dominant mean structure with kriging for 
estimating the random effect.
The mean of RIMSE obtained by PDE+ is 48.780, which is about 13\% $(=55.081/48.780 -1)$ improvement over kriging, in addition to the interpretable 
predictive model.

\subsection{German PM$_{10}$ Data}
We applied the proposed method to a dataset of air quality obtained from European Commission's Airbase and air quality e-reporting repositories (AQER).
The monthly averages of PM$_{10}$ concentrations within Germany between January 2016 and June 2018 are considered.
The package \texttt{saqgetr} on CRAN was used to extract hourly PM$_{10}$ averages, and an aggregation to obtain monthly averages was performed 
at each of background stations.
Totally 221 stations have complete monthly records over the 30 months considered.
The monthly average concentrations  for 221 stations shown in Figure 4(A)  indicate several peaks occurring particularly at the beginning of years.
The map of station locations is shown in Figure 4(B).
For illustration, we first standardize $\x(\s)$ for being centered and removing scale for further analysis.
We randomly partition $\{\y(\s_i,\cdot); i=1,\ldots,221\}$ into a learning data with size 177 and a testing data with size 44.
In one single analysis run, we conduct the eigenvalue decomposition in (\ref{mrphd}) for the learning data to obtain initial estimates $\hat \bt_j$'s.
Two significant directions are found by pe-mrPHD with the eigenvalues (1.79, 0.71, 0.59, 0.20).
After the convergence of PDE algorithm, two leading directions, (-0.321, -0.196, 0.389, 0.841)$'$ and (0.524, 0.763, 0.285, 0.246)$'$, for $\x(\s)$ 
variates are obtained by choosing the bandwidths $h_Y=6$ and $h_\x=1$.
It reveals that there exists respective dominant effects; namely, {\it squared latitude} for the first variate, and a linear combination of  {\it longitude} 
and {\it latitude} for the second variate.

Recall that both $w_j{(t)}$ and $f_j(\s)$ affect $z(\s,t)$ through the inner product form.
Figure 4(C) displays the first estimated basis function with a conceivable \lq \lq W'' pattern reflecting the peaks of the data.
Interestingly, $\hat{f}_1$ shown in Figure 4(E) has a bend down pattern with larger values  occurring at $\hat \bt_1'\x(\s)$ around zero.
That implies larger positive (smaller negative) values of $\hat \bt_1'\x(\s)$  indicate locations of northern (southern) regions due to the dominant 
squared latitude  effect in $\hat \bt_1$.
As a result, the influence of the first dominant component $\hat{w}_1(t)\hat{f}_1(\s)$ on $z(\s,t)$ shows that air pollution of PM$_{10}$ in Germany 
is usually heavier in the mid-latitude areas; moreover, it is more serious in spring and winter than in summer and autumn.
Due to the {\it longitude} and  {\it latitude} effect in $\hat \bt_2$, roughly standing for equally weight from northeast to southwest; specifically, larger 
positive (smaller negative) values of $\hat \bt_2'\x(\s)$  indicate locations of northeast (southwest) regions.
Figure 4(F) shows a linearly decreasing pattern along $\hat \bt_2' \x(\s)$.
Figure 4(D) displays a symmetric pattern fluctuating around zero in the vertical axis.
From the magnitudes compared to Figure 4(E), the function of $\hat{w}_2(t)\hat{f}_2(\s)$ may serve as the complement component of influence on 
PM$_{10}$.

The results based on 100 random data partitions as previously mentioned for sampling performance are summarized in Table 4.
The PDE indeed captures the interesting functional patterns for $w_j(t)$ and $f_j(\s)$, $j=1,2$.
In the aspect of predictive accuracy, PDE+ does outperform with the smallest averaged prediction error, as shown in Table 4, even though kriging 
only  is inferior to spTimer.
However, mgcv is prone to overfitting.
Without using smoother functions in (\ref{mgcv}) (i.e. restricting the wiggliness of $q$'s instead of dropping the $q_7$ term), mgcv produces a mean 
of 193.28 with a standard error 58.49 for RIMSE and a mean of 50.98 with a standard error 21.15 for RPMSE.
In contrast, the mean of RIMSE obtained by PDE+ is only 16.658, which is about 70\% $(=28.462/16.658 -1)$ improvement over spTimer.
More importantly, PDE+ can greatly help the explanation of the spatio-temporal trends, which is difficult for all other methods.

\section{Discussion}
Effective dimension reduction is much challenging statistical issue on spatio-temporal data, which was less paid attention in the past decades.
We first focus on dimension reduction for spatial covariates $\x(\s)$.
Searching optimum of possible linear combinations for $\x(\s)$ is crucial for solving such problem.
Another important issue is finding the connection of linear or nonlinear association with the 
spatio-temporal variable of interest.
The PDE+ can successfully achieve the above goals.
The proposed method  based on dimension reduction effectively demonstrates accurate prediction from the results of numerical examples 
and applications.
In contrast, few methods utilize a data-driven approach for providing mean structures in the literature.

One future topic of research is how to extend the proposed method to  spatio-temporal covariates.
Missing data often occur in practical problems, which is another research topic requiring more investigation.

\vskip.3in \noindent
\section*{REFERENCES}

Bakar, K. S., and Sahu, S. K. (2015). \lq \lq  spTimer: Spatio-temporal bayesian modelling using R,'' Journal of Statistical Software, 63
, 1-32.

Banerjee, S., Gelfand, A. E., Finley, A. O., and Sang, H. (2008). \lq \lq Gaussian predictive process models for large spatial data sets,'' Journal of the 
Royal Statistical Society: Series B (Statistical Methodology), 70
, 825-848.

Braud, I., and Obled, C. (1991),  \lq \lq On the use of Empirical Orthogonal Function (EOF) analysis in the simulation of random fields,'' Stochastic 
Hydrology and Hydraulics, 5
, 125-134.

Coudret, R., Girard, S., and Saracco, J. (2014), \lq \lq A new sliced inverse regression method for multivariate response,'' Computational Statistics 
\& Data Analysis, 77, 285-299.

Cressie, N. (2015). Statistics for spatial data. John Wiley \& Sons.

Cressie, N., and Johannesson, G. (2008), \lq \lq Fixed rank kriging for very large spatial data sets,'' Journal of the Royal Statistical Society: Series 
B, 70
, 209-226.

Cressie, N., Shi, T., and Kang, E. L. (2010), \lq \lq Fixed rank filtering for spatio-temporal data,'' Journal of Computational and Graphical Statistics, 19
, 724-745.

Cressie, N. and Wikle, C. K. (2011),  Statistics for Spatio-Temporal Data, Hoboken, NJ: John Wiley \& Sons.


Demsar, U., Harris, P., Brunsdon, C., Fotheringham, A. S., and McLoone, S. (2013),  \lq\lq Principal component analysis on spatial data: an overview,'' Annals of the Association of American Geographers, 103
, 106-128.


Fass\`o, A., and Cameletti, M. (2010), \lq\lq A unified statistical approach for simulation, modeling, analysis and mapping of environmental data.'' Simulation, 86
, 139-153.



Lee, D. J., and Durb\'an, M. (2011). \lq\lq P-spline ANOVA-type interaction models for spatio-temporal smoothing,'' Statistical modelling, 11
, 49-69.

Li, K. C., Aragon, Y., Shedden, K., and Agnan, C. T. (2003), \lq\lq Dimension reduction for multivariate response data,'' Journal of the American Statistical 
Association, 98
, 99-109.

Lue, H. H. (2009), \lq\lq Sliced inverse regression for multivariate response regression,'' Journal of Statistical Planning and Inference, 139, 2656-2664.

Lue, H. H. (2010), \lq\lq On principal Hessian directions for multivariate response regressions," Computational Statistics, 25, 619-632.

Lue, H. H. (2019). \lq\lq Pairwise directions estimation for multivariate response regression data.'' Journal of Statistical Computation and Simulation, 89
, 776-794.


Martin, J. D., and Simpson, T. W. (2005), \lq\lq Use of kriging models to approximate deterministic computer models,'' AIAA journal, 43, 853-863.

Pebesma, E., and  Heuvelink, G. (2016). \lq \lq Spatio-temporal interpolation using gstat,'' RFID Journal, 8
, 204-218.

Sharples, J. J., and Hutchinson, M. F. (2005). \lq\lq Spatio-temporal analysis of climatic data using additive regression splines,'' In Proceedings of International 
Congress on Modelling and Simulation (MODSIM'05), pp. 1695-1701.

Sherman, M. (2011). Spatial statistics and spatio-temporal data: covariance functions and directional properties. John Wiley \& Sons.

Thorson, J. T., Cheng, W., Hermann, A. J., Ianelli, J. N., Litzow, M. A., O'Leary, C. A., and Thompson, G. G. (2020), \lq\lq Empirical orthogonal function regression: 
Linking population biology to spatial varying environmental conditions using climate projections,''  Global Change Biology.

Tzeng, S., and Huang, H. C. (2018), \lq\lq   Resolution adaptive fixed rank kriging,'' Technometrics, 60, 198-208.

Wang, W. T., and Huang, H. C. (2017).  \lq\lq  Regularized principal component analysis for spatial data,'' Journal of Computational and Graphical Statistics, 26
, 14-25.

Wikle, C. K. (2010). \lq\lq Low-rank representations for spatial processes.'' in Handbook of spatial statistics, pp. 107-118.

Wood, S. N. (2017). Generalized additive models: an introduction with R. CRC press.

Xia, Y., Tong, H., Li, W. K., and Zhu, L. X. (2002), \lq\lq An adaptive estimation of dimension reduction space,'' Journal of the Royal Statistical Society: Series B, 
64, 363-410.

\vfil
\eject

\begin{figure}
\begin{centering}
\begin{tabular}{c}
\includegraphics[scale=0.98, trim= 5cm 4cm 0.5cm 5cm, clip=true]{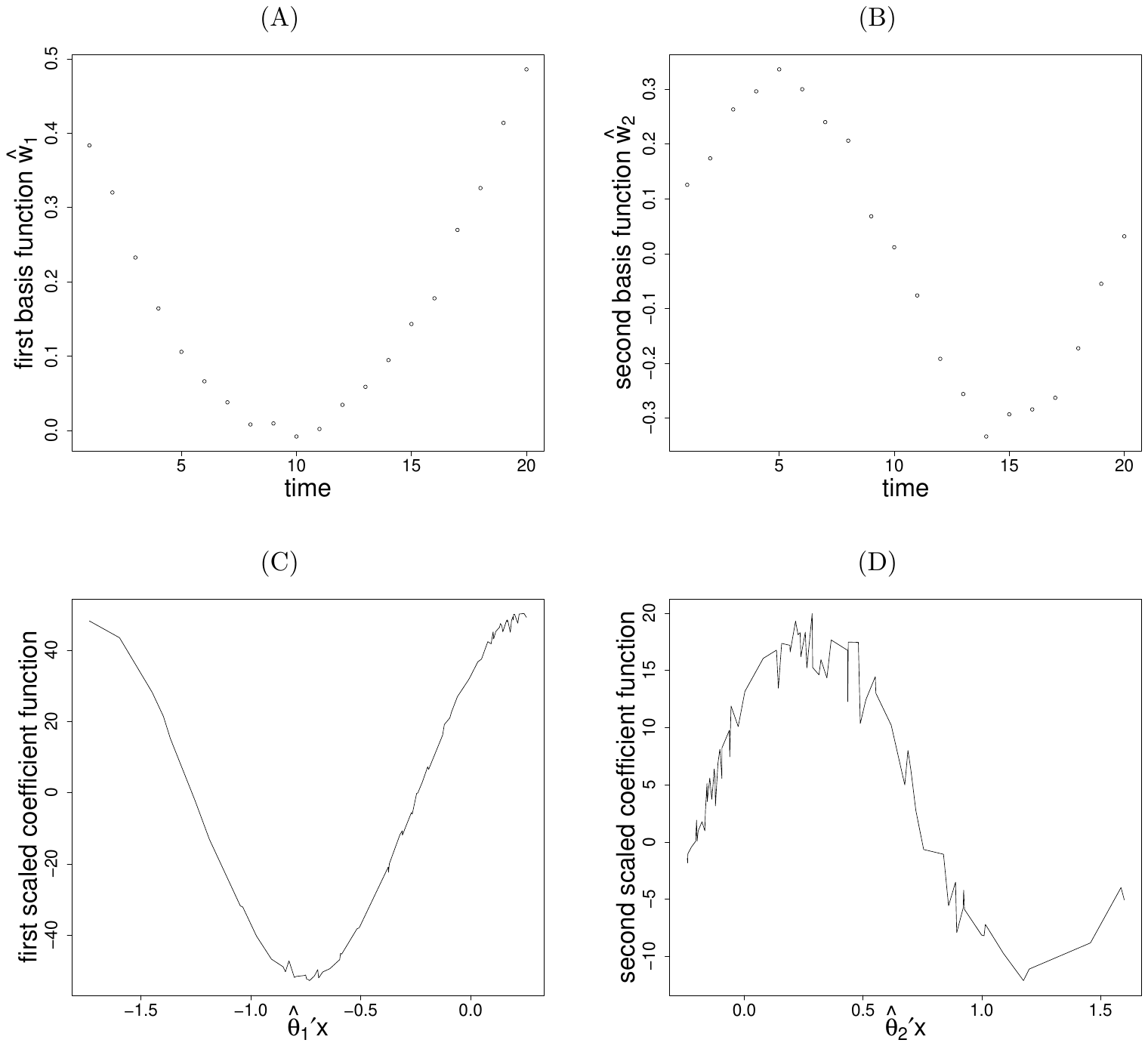}
\end{tabular}
\par\end{centering}
\caption{Plots of (A) the first and (B) the second basis functions with (C) the first and (D) the second scaled coefficient functions for a 
randomly selected simulation from model in (\ref{ex1a}) and (\ref{ex1b}).}
\end{figure}

\vfil
\eject

\begin{figure}
\begin{centering}
\begin{tabular}{c}
\includegraphics[scale=0.98, trim= 5cm 4cm 0.5cm 5cm, clip=true]{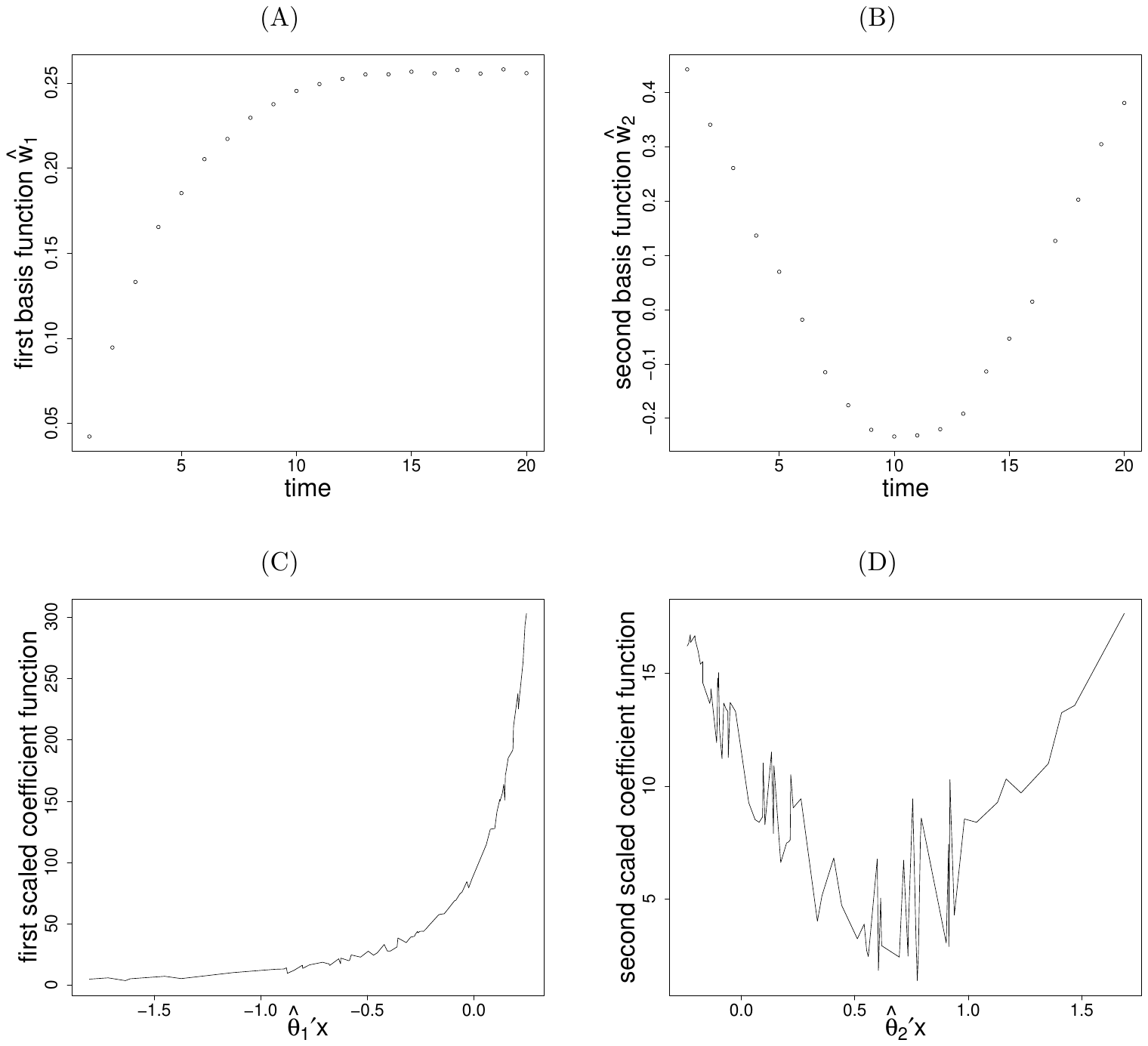}
\end{tabular}
\par\end{centering}
\caption{Plots of (A) the first and (B) the second basis functions with (C) the first and (D) the second scaled coefficient functions for a 
randomly selected simulation from model in (\ref{ex2a}) and (\ref{ex2b}).}
\end{figure}

\vfil
\eject

\begin{figure}
\begin{centering}
\begin{tabular}{c}
\includegraphics[scale=0.98, trim= 5cm 2.2cm 1cm 4.4cm, clip=true]{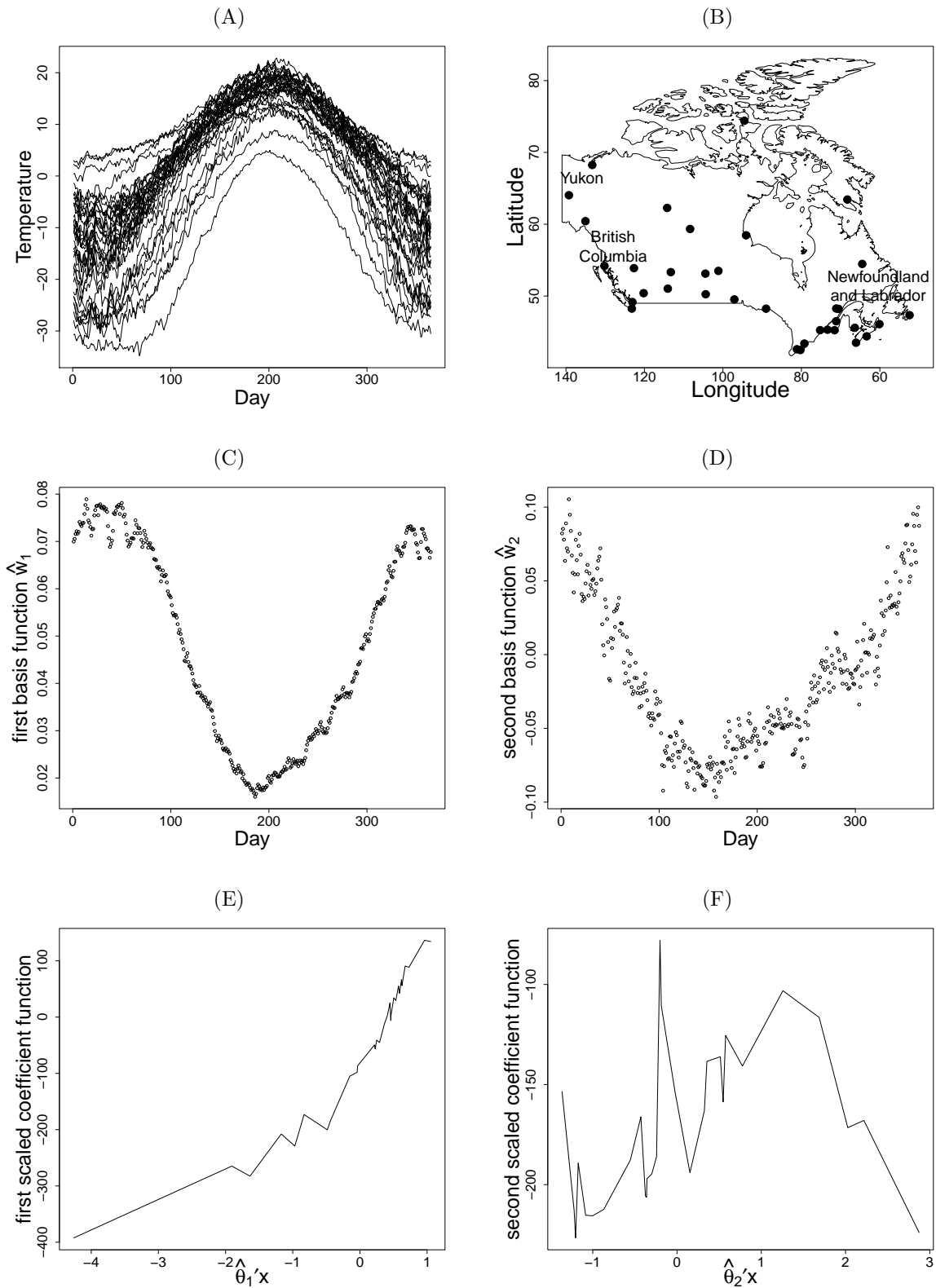} 
\end{tabular}
\par\end{centering}
\caption{Plots of (A) the average daily temperatures over individual locations, (B) the spatial distribution of locations, (C) the first and 
(D) the second basis functions, and (E) the first and (F) the second scaled coefficient functions for Canadian Weather data.}
\end{figure}

\vfil
\eject

\begin{figure}
\begin{centering}
\begin{tabular}{c}
\includegraphics[scale=0.97, trim= 3cm 3.8cm 1cm 4cm, clip=true]{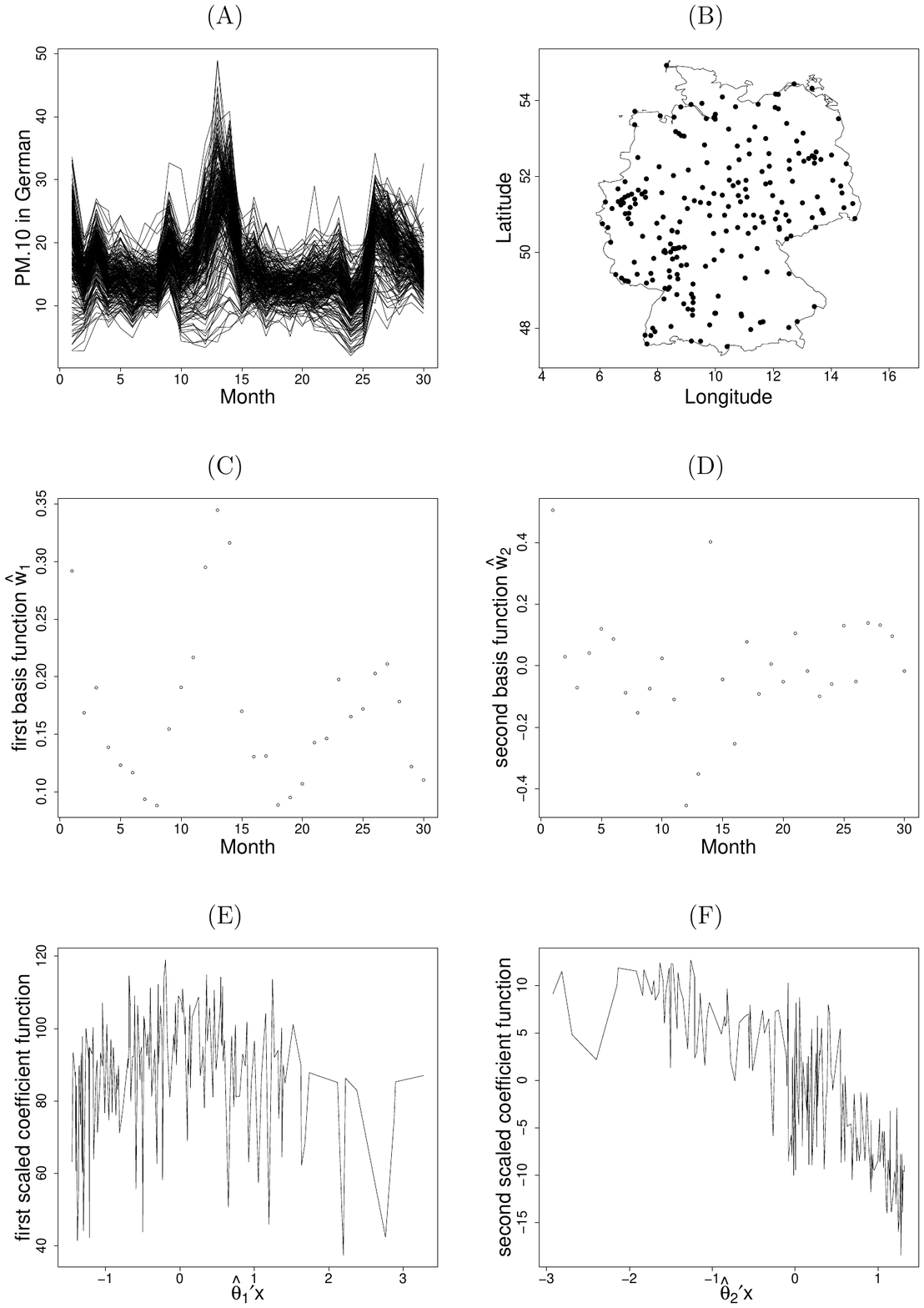} 
\end{tabular}
\par\end{centering}
\caption{Plots of (A) the monthly averages of PM$_{10}$ concentrations over individual locations, (B) the spatial distribution of locations, (C) the first 
and (D) the second basis functions, and (E) the first and (F) the second scaled coefficient functions for German PM$_{10}$ data.}
\end{figure}

\vfil
\eject

\centerline{Table 1. \hskip.05in Averaged RIMSE and RPMSE from different methods  for data generated from model }
\centerline{ in (\ref{ex1a}) and (\ref{ex1b}) in 100 replicates (values given in parentheses are the corresponding standard errors).}
$$\vbox{\tabskip=0pt
  \offinterlineskip\halign {\strut#& 
  #\tabskip=.10em plus .10em&\hfil #\hfil&
  #\tabskip=.10em plus .10em&\hfil #\hfil&
  #\tabskip=.10em plus .10em&\hfil #\hfil&
  #\tabskip=.10em plus .10em&\hfil #\hfil&
  #\tabskip=.10em plus .10em&\hfil #\hfil&
  #\tabskip=.10em plus .10em&\hfil #\hfil&
  #\tabskip=.10em plus .10em&\hfil #\hfil&
  #\tabskip=0pt\cr
\noalign{\hrule} 
&&    &&    PDE+  &&  naive  &&  SpTimer  &&  kriging  &&  mgcv  && \cr
\noalign{\hrule}
&&    \textrm{RIMSE}    &&    4.517   &&  37.103  &&  35.354   &&   5.912   &&  6.598    && \cr  
&&                                 &&   (1.119)  &&  (3.412)  &&  (3.183)  &&  (1.197)  && (1.015)  && \cr 
&&    \textrm{RPMSE}   &&    1.188   &&   8.902   &&   8.480    &&   1.618   &&  1.638   && \cr
&&                                 &&   (0.460) &&   (0.641)  &&  (0.611)  && (0.433)   && (0.380)  && \cr
\noalign{\hrule} 
\cr}}$$

\vskip.5in
\centerline{Table 2. \hskip.05in Averaged RIMSE and RPMSE from different methods  for data generated from model }
\centerline{ in (\ref{ex2a}) and (\ref{ex2b}) in 100 replicates (values given in parentheses are the corresponding standard errors).}
$$\vbox{\tabskip=0pt
  \offinterlineskip\halign {\strut#& 
  #\tabskip=.10em plus .10em&\hfil #\hfil&
  #\tabskip=.10em plus .10em&\hfil #\hfil&
  #\tabskip=.10em plus .10em&\hfil #\hfil&
  #\tabskip=.10em plus .10em&\hfil #\hfil&
  #\tabskip=.10em plus .10em&\hfil #\hfil&
  #\tabskip=.10em plus .10em&\hfil #\hfil&
  #\tabskip=.10em plus .10em&\hfil #\hfil&
  #\tabskip=0pt\cr
\noalign{\hrule} 
&&    &&    PDE+  &&  naive  &&  SpTimer  &&  kriging  &&  mgcv  && \cr
\noalign{\hrule}
&&    \textrm{RIMSE}    &&   5.759   &&   65.722   &&  60.496    &&   7.375   &&  13.375    && \cr  
&&                                 &&  (1.313)  &&   (8.934)  &&   (8.224)   &&  (2.171)  &&  (2.581)   && \cr 
&&    \textrm{RPMSE}   &&   1.654   &&  17.656    &&  16.316    &&   2.612   &&    4.056    && \cr
&&                                 &&  (0.545)  &&   (2.602)  &&   (2.506)   &&  (0.988)  &&   (0.731)  && \cr
\noalign{\hrule} 
\cr}}$$

\vfil

\centerline{Table 3. \hskip.05in Averaged RIMSE and RPMSE from different methods for Canadian Weather data}
\centerline{in 100 validations (values given in parentheses are the corresponding standard errors).}
$$\vbox{\tabskip=0pt
  \offinterlineskip\halign {\strut#& 
  #\tabskip=.10em plus .10em&\hfil #\hfil&
  #\tabskip=.10em plus .10em&\hfil #\hfil&
  #\tabskip=.10em plus .10em&\hfil #\hfil&
  #\tabskip=.10em plus .10em&\hfil #\hfil&
  #\tabskip=.10em plus .10em&\hfil #\hfil&
  #\tabskip=.10em plus .10em&\hfil #\hfil&
  #\tabskip=.10em plus .10em&\hfil #\hfil&
  #\tabskip=0pt\cr
\noalign{\hrule} 
&&    &&    PDE+  &&  naive  &&  SpTimer  &&  kriging  &&  mgcv  && \cr
\noalign{\hrule}
&&    \textrm{RIMSE}    &&  48.780   &&  112.874  &&  107.558   &&   55.081  &&   73.51   && \cr  
&&                                 &&  (26.03)  &&   (32.60)   &&   (32.52)   &&   (51.69)  &&  (72.60)  && \cr 
&&     \textrm{RPMSE}  &&    2.969   &&     6.685   &&     6.436    &&    3.318   &&    5.128  && \cr
&&                                 &&   (2.03)   &&     (2.04)   &&    (2.03)    &&    (2.74)   &&   (6.47)   && \cr
\noalign{\hrule} 
\cr}}$$

\vskip.5in
\centerline{Table 4. \hskip.05in Averaged RIMSE and RPMSE from different methods for German PM$_{10}$ data}
\centerline{in 100 validations (values given in parentheses are the corresponding standard errors).}
$$\vbox{\tabskip=0pt
  \offinterlineskip\halign {\strut#& 
  #\tabskip=.10em plus .10em&\hfil #\hfil&
  #\tabskip=.10em plus .10em&\hfil #\hfil&
  #\tabskip=.10em plus .10em&\hfil #\hfil&
  #\tabskip=.10em plus .10em&\hfil #\hfil&
  #\tabskip=.10em plus .10em&\hfil #\hfil&
  #\tabskip=.10em plus .10em&\hfil #\hfil&
  #\tabskip=.10em plus .10em&\hfil #\hfil&
  #\tabskip=0pt\cr
\noalign{\hrule} 
&&    &&    PDE+  &&  naive  &&  SpTimer  &&  kriging  &&  mgcv  && \cr
\noalign{\hrule}
&&    \textrm{RIMSE}    &&  16.658    &&  45.905   &&  28.462    &&   37.167   &&   56.452   && \cr  
&&                                 &&   (1.221)  &&   (0.704)  &&   (0.824)   &&   (1.445)  &&    (3.778)  && \cr 
&&    \textrm{RPMSE}   &&   3.479    &&    8.440    &&    5.333    &&    6.755    &&   10.783   && \cr
&&                                 &&  (0.290)   &&   (0.140)   &&   (0.166)   &&  (0.437)   &&   (0.675)   && \cr
\noalign{\hrule} 
\cr}}$$

\end{document}